\documentclass[%
11pt,tightenlines,
onecolumn,aip,jmp,amsmath,amssymb,eqsecnum,amsfonts,
nofootinbib,reprint,%
]{revtex4-2}

\usepackage{graphicx}
\usepackage{dcolumn}
\usepackage{bm}

\begin{document}

\title{On the supersymmetry of the Klein-Gordon oscillator\\[2mm] \normalsize{Dedicated to Akira Inomata on the occasion of his 90$^{\bf th}$ birthday\\~ }\\ }

\author{Georg Junker}\email{gjunker@eso.org; georg.junker@fau.de}
\affiliation{European Organization for Astronomical Research in the Southern Hemisphere,
Karl-Schwarzschild-Stra{\ss}e 2, D-85748 Garching, Germany\\
{\rm and}\\
Institut f\"ur Theoretische Physik I, Universit\"at Erlangen-N\"urnberg, Staudtstra{\ss}e 7, D-91058 Erlangen, Germany;
}%
\date{\today}

\begin{abstract}
The three-dimensional Klein-Gordon oscillator is shown to exhibit an algebraic structure known from supersymmetric quantum mechanics. The supersymmetry is found to be unbroken with a vanishing Witten index, and it is utilized to derive the spectral properties of the Klein-Gordon oscillator, which is closely related to that of the non-relativistic harmonic oscillator in three dimensions. Supersymmetry also enables us to derive a closed-form expression for the energy-dependent Green's function.
\end{abstract}
\maketitle

\newcommand{\rmi}{\textrm{i}}
\newcommand{\rmd}{\textrm{d}}
\newcommand{\balpha}{\bm{\alpha}}
\newcommand{\bsigma}{\bm{\sigma}}
\newcommand{\rme}{\textrm{e}}
\newcommand{\JPA}{J.\ Phys.\ A}
\newcommand{\RMP}{Rev.\ Mod.\ Phys.\ }
\newcommand{\PRL}{Phys.\ Rev.\ Lett.\ }
\newcommand{\JMP}{J.\ Math.\ Phys.\ }
\renewcommand*{\thefootnote}{\fnsymbol{footnote}}

\section{Introduction}\label{Section1}
Starting with Galileo's pendulum experiment \cite{Palmieri2009} in 1602 and with Hook's law of elasticity \cite{Hook1678} from 1678, harmonic oscillators played significant roles in classical physics. More importantly, the harmonic oscillator was the first system to which early quantum theory was successfully applied by Planck \cite{Planck1900} in 1900 when developing his law of black body radiation. Nowadays the harmonic oscillator is a standard part of any introductory text book on non-relativistic quantum mechanics. In relativistic quantum mechanics the harmonic oscillator was initially studied within Dirac's theory of electrons in the 1960s \cite{Ito1965,Swamy1969,Cook1971} but  attracted considerable attention only with the seminal work by Moshinsky and Szczepaniak \cite{Moshinsky1989}, see also Quesne and Moshinsky \cite{Quesne1990}. Inspired by this so-called Dirac oscillator, the Klein-Gordon oscillator (KGO) has been studied by various authors \cite{Debergh1992,Bruce1993,Dvoeglazov1995}.

The KGO Hamiltonian characterises a relativistic spin-zero particle with mass $m$ minimally coupled to a complex linear vector potential. Since its introduction, the KGO has and still is attracting interest.
The spectral properties of the one-dimensional system were discussed, for example, in \cite{Rao2008,Boumali2011}. For a treatment in non-commutative space see the work \cite{Mirza2004,Mirza2011}, for recent results in a non-trivial topology see \cite{Santos2019,Ahmed2020,Braganca2020,Zhong2021} and references therein.

At least since 1990 it is know that the Dirac oscillator exhibits a supersymmetric (SUSY) structure, which in turn allows for explicit solutions \cite{Benitez1990,Beckers1990,Quesne1991,Romero1991}. More recently, SUSY also enabled us to formulate Feynman's path integral approach for Dirac systems \cite{Junker2018}. Let us emphasize that SUSY in the current context is not based on the original idea which transforms between states with different internal spin-degree of freedom. SUSY here refers to what nowadays is commonly known
as supersymmetric quantum mechanics, see for example \cite{Junker2019} and reference therein.

The purpose of the present work is two-fold. First we show that the Klein-Gordon oscillator possesses a hidden SUSY in the aforementioned sense. Secondly we will derive an explicit expression for the Green function of the KGO. In doing so we will closely follow the generic approach for SUSY in relativistic Hamiltonians with fixed but arbitrary spin \cite{Junker2020}.

In the next section we will set up the stage with a brief discussion on the KGO Hamiltonian in three space dimensions and show that this Hamiltonian exhibits a SUSY structure by mapping it onto a SUSY quantum mechanical system. This fact is then utilized to derive explicit results of the system. That is, in section 3 we will derive the eigenvalues and associated eigenstates, followed by section 4 where we derive the corresponding Green's function in a closed form. Finally, section 5 closes with a summary and some comments.

\section{Supersymmetry of the KGO}\label{Section2}
The Hamiltonian form of the Klein-Gordon equation with an arbitrary vector potential was originally introduced by Feshbach and Villars \cite{Feshbach1958} from which the KGO Hamiltonian may be constructed via the minimal coupling $\vec{p}\to\vec{\pi}:= \vec{p}-\rmi m\omega\vec{r}$, where $m>0$ stands for the mass of the spin-less Klein-Gordon particle and $\omega >0$ is a coupling constant to be identified with the harmonic oscillator frequency.

This minimal coupling might be interpreted as a complex-valued vector potential of the form $\vec{A}(\vec{r}):= \rmi (mc\omega/q)\vec{r}$ with $q$ being the charge of the particle and $c$ the speed of light. However, such a vector potential is not linked to any kind of gauge invariance as $\vec{A}(\vec{r})=\rmi (mc\omega/2q)\vec{\nabla}r^2$ cannot be gauged away by a pure phase-factor in the wave function due to the presence of the imaginary unit.

First explicit expressions of the KGO Hamiltonian were presented by Debergh et al \cite{Debergh1992} for an isotropic system. The KGO Hamiltonian for a more general anisotropic oscillator system is due to Bruce and Minning \cite{Bruce1993}. For the sake of simplicity we will consider the isotropic system characterised by the Hamiltonian
\begin{equation}\label{HKG}
  {\cal H} := \frac{\vec{\pi}^\dag\cdot\vec{\pi}}{2m}\otimes\left(\tau_3+\rmi \tau_2\right)+mc^2\otimes\tau_3
\end{equation}
acting on the Hilbert space $L^2(\mathbb{R}^3)\otimes\mathbb{C}^2$. In the above the $\tau_i$'s stand for the Pauli matrices
\begin{equation}\label{taus}
  \tau_1 := \left( \begin{array}{cc}  0 & 1 \\ 1 & 0 \end{array}\right)\,,\qquad
  \rmi\tau_2 := \left( \begin{array}{cc}  0 & 1 \\ -1 & 0 \end{array}\right)\,,\qquad
  \tau_3 := \left( \begin{array}{cc}  1 & 0 \\ 0 & -1 \end{array}\right)\,.
\end{equation}
Note that these Pauli matrices do not represent a spin-degree of freedom. The 2-spinors on which above Hamiltonian acts are those originally introduced by Feshbach and Villars \cite{Feshbach1958}.

The KGO Hamiltonian (\ref{HKG}) is pseudo-Hermitian \cite{Mostafazadeh2003,Mostafazadeh2006}, that is ${\cal H}^\dag = \tau_3\,{\cal H}\,\tau_3$, and reads in an explicit $2\times 2$ matrix notation
\begin{equation}\label{HKG2}
  {\cal H}=\left(\begin{array}{cc}
                   M & A \\
                   -A & -M
                 \end{array}
  \right)\,,
\end{equation}
where we have set $M:= H_{\rm NR}+mc^2$ and $A := H_{\rm NR}$, both being in essence represented by the non-relativistic harmonic oscillator Hamiltonian in three dimensions
\begin{equation}\label{HNR}
  H_{\rm NR} := \frac{\vec{\pi}^\dag\cdot\vec{\pi}}{2m} = \frac{1}{2m}\vec{p}\,^2 +\frac{m}{2}\omega^2\vec{r}\,^2 -\frac{3}{2}\hbar\omega\,.
\end{equation}
Here and in the following we will use calligraphic symbols for operators acting on the full Hilbert space $L^2(\mathbb{R}^3)\otimes\mathbb{C}^2$ and operators represented in italic act on the subspace $L^2(\mathbb{R}^3)$.

Obviously, the diagonal and off-diagonal elements in (\ref{HKG2}) commute, i.e., $[M,A]=0$. Hence, following the general approach of \cite{Junker2020} it is possible to establish an $N=2$ SUSY structure as follows.
\begin{equation}\label{HQSUSY}
  \begin{array}{l}
    {\cal H}_{\rm SUSY} := \displaystyle\frac{1}{2mc^2}\left({\cal M}^2-{\cal H}^2\right)= \frac{1}{2mc^2}H^2_{\rm NR}\otimes 1\,, \\[4mm]
    {\cal Q}:= \displaystyle\frac{1}{\sqrt{2mc^2}} \left(\begin{array}{cc} 0 & A \\ 0 & 0 \end{array}\right)\,,\qquad
    {\cal W}:= \left(\begin{array}{cc} 1 & 0 \\ 0 & -1 \end{array}\right)\equiv \tau_3\,,
  \end{array}
\end{equation}
where we have set ${\cal M} := M\otimes 1$. The above SUSY operators obey the SUSY algebra
\begin{equation}\label{SUSY}
\begin{array}{c}
  {\cal H}_{\rm SUSY}=\{{\cal Q},{\cal Q}^\dag\}\,,\quad {\cal Q}^2 =0={{\cal Q}^\dag}^2 \,,\\[2mm]
  [{\cal W}, {\cal H}_{\rm SUSY}]=0 \,, \quad \{{\cal Q},{\cal W}\}=0=\{{\cal Q}^\dag,{\cal W}\}\,.
\end{array}
\end{equation}
Let us note that in the current context the third Pauli matrix plays the role of the Witten party operator ${\cal W}$. Therefore, the upper and lower components of a general 2-spinor belong to the subspace with positive and negative Witten parity, respectively. We further
remark that $\dim\ker {\cal Q}=\dim\ker {\cal Q}^\dag = \dim\ker H_{\rm NR}=1$. That is, SUSY is unbroken as ${\cal H}_{\rm SUSY}$ has zero-energy eigenstates \cite{Junker2019}, but the Witten index $\Delta$ still vanishes as
\begin{equation}\label{Delta}
  \Delta := {\rm ind\,} {\cal Q} = \dim\ker {\cal Q}- \dim\ker {\cal Q}^\dag = 0\,.
\end{equation}
To the best of our knowledge this is the first quantum mechanical system with an unbroken $N=2$ SUSY but vanishing Witten index, implying that the spectrum of ${\cal H}$ is fully symmetric with respect to the origin as we will see in the following section.

\section{Spectral properties of the KGO}\label{Section3}
As has been shown recently \cite{Junker2020}, SUSY in a relativistic Hamiltonian implies the existence of a Foldy-Wouthuysen transformation which brings that Hamiltonian into a block-diagonal form. In the case of the KGO this transformation operator ${\cal U}$, which is a pseudo-unitary operator in the sense that ${\cal U}^{-1}=\tau_3 \, {\cal U}^\dag\tau_3$, reads
\begin{equation}\label{U}
  {\cal U}:=\frac{|{\cal H}|+\tau_3{\cal H}}{\sqrt{2({\cal H}^2+{\cal M}|{\cal H}|)}}
\end{equation}
leading to the block-diagonal Foldy-Wouthuysen Hamiltonian
\begin{equation}\label{HFW}
  {\cal H}_{\rm FW}:= {\cal U}\,{\cal H}\,{\cal U}^{-1} =H_{\rm FW}\otimes \tau_3\,,\qquad H_{\rm FW}:=\sqrt{2mc^2H_{\rm NR}+m^2c^4}\,.
\end{equation}
To be a bit more explicit,
let us define $\tanh\Theta := A/M = H_{\rm NR}/(H_{\rm NR} +mc^2)$, which then allows us to write the transformation (\ref{U}) in the matrix form \cite{Debergh1992}
\begin{equation}\label{U2x2}
   {\cal U} = \left(
   \begin{array}{cc}
     \cosh\frac{\Theta}{2} & \sinh\frac{\Theta}{2} \\[2mm] \sinh\frac{\Theta}{2} & \cosh\frac{\Theta}{2}
   \end{array} \right)=
   \frac{1}{\sqrt{2}}\left(\begin{array}{cc}
     \sqrt{\frac{M}{H_{\rm FW}}+1} & \sqrt{\frac{M}{H_{\rm FW}}-1} \\[2mm]
     \sqrt{\frac{M}{H_{\rm FW}}-1} & \sqrt{\frac{M}{H_{\rm FW}}+1}
   \end{array} \right)\,.
\end{equation}
Above expressions are functions of operators all of which may be expressed in terms of $H_{\rm NR}$. Hence, using the spectral theorem these are well-defined. In fact,
with the spectral properties of the non-relativistic harmonic-oscillator Hamiltonian (\ref{HNR}) one can directly obtain those of (\ref{HKG}). Let $\psi_{n\ell \mu}$ denote the well-known eigenfunctions of $H_{\rm NR}$ corresponding to the eigenvalue $\varepsilon_{n\ell}$, then we have
\begin{equation}\label{HNEphi}
  \begin{array}{l}
    H_{\rm NR}\,\psi_{n\ell \mu} = \varepsilon_{n\ell}\,\psi_{n\ell \mu}\,,\qquad \varepsilon_{n\ell} = \hbar\omega(2n+\ell)\,,\qquad n,\ell\in\mathbb{N}_0\,, \\[2mm]
    \displaystyle
    \psi_{n\ell \mu}(\vec{r})= \left(\frac{m\omega}{\hbar}\right)^{\ell/2 + 3/4}\sqrt{\frac{2 \, n!}{\Gamma(n+\ell +3/2)}}\, r^\ell\,
                             \rme^{-m\omega r^2/\hbar}\, {\rm L}^{\ell +1/2}_n\left(\frac{m\omega}{\hbar}\,r^2\right){\rm Y}_{\ell\mu}(\vec{e})\,,\\[4mm]
    \mu \in \{ -\ell,-\ell + 1,\ldots,\ell - 1, \ell\}\,,\qquad r:=|\vec{r}|\,, \qquad\vec{e}:= \vec{r}/r\,,
  \end{array}
\end{equation}
where ${\rm L}^{\ell +1/2}_n$ and ${\rm Y}_{\ell\mu}$ denote the associated Laguerre polynomials and the spherical harmonics, respectively. See, for example, ref.\ \cite{GP1990}.
Then, the eigenvalues and eigenfunctions of (\ref{HKG}) are explicitly given by
\begin{equation}\label{HPsi}
\begin{array}{l}
  \displaystyle
  {\cal H}\Psi^\pm_{n\ell\mu}=E^\pm_{n\ell}\Psi^\pm_{n\ell\mu}\,,\qquad E_{n\ell}^\pm=\pm mc^2\sqrt{1+\frac{2\varepsilon_{n\ell}}{mc^2}}\,,\\[4mm]
  \Psi^+_{n\ell\mu}(\vec{r})=\psi_{n\ell\mu}(\vec{r})\left(\begin{array}{r}\cosh\frac{\vartheta_{n\ell}}{2} \\[1mm] -\sinh\frac{\vartheta_{n\ell}}{2}\end{array}\right)\,,\qquad
  \Psi^-_{n\ell\mu}(\vec{r})=\psi_{n\ell\mu}(\vec{r})\left(\begin{array}{r}-\sinh\frac{\vartheta_{n\ell}}{2} \\[1mm] \cosh\frac{\vartheta_{n\ell}}{2}\end{array}\right)\,,
\end{array}\end{equation}
where $\tanh\vartheta_{n\ell} := \varepsilon_{n\ell}/(\varepsilon_{n\ell}+mc^2)$.
These states form an orthonormal basis in $L^2(\mathbb{R}^2)\otimes\mathbb{C}^2$ with respect to the scalar product \cite{Feshbach1958}.
\begin{equation}\label{SP1}
  \langle\Psi_1|\Psi_2\rangle := \int_{\mathbb{R}^3}\rmd^3\vec{r}\,\, \overline{\Psi}_1(\vec{r})\,\tau_3\,\Psi_2(\vec{r})\,,
\end{equation}
where the overbar stands for the transposed and complex conjugated $2$-spinor.
That is,
\begin{equation}\label{SP2}
  \langle\Psi^\pm_{n\ell\mu}|\Psi^\pm_{n'\ell'\mu'}\rangle =\pm \delta_{nn'}\delta_{\ell\ell'}\delta_{\mu\mu'}\,,\qquad
  \langle\Psi^\pm_{n\ell\mu}|\Psi^\mp_{n'\ell'\mu'}\rangle = 0\,.
\end{equation}
Obviously, the scalar product (\ref{SP1}), which was already introduced by Feshbach and Villars \cite{Feshbach1958}, is not positive definite and might raise some questions on its probabilistic interpretation. However, Mostafazadeh's theory of pseudo-Hermitian operators does provide a solution to this obstacle. The Klein-Gordon case was explicitly discussed in \cite{Mostafazadeh2003,Mostafazadeh2006}.

Let us conclude this section by noting that the SUSY ground states associated with the non-relativistic eigenvalue $\varepsilon_{00}=0$ are given by
\begin{equation}\label{Psi00}
  \Psi^+_{000}(\vec{r})=\left(\frac{m\omega}{\hbar\pi}\right)^{3/4}\rme^{-m\omega r^2/\hbar}\left(\begin{array}{c}1 \\ 0\end{array}\right)\,,\qquad
  \Psi^-_{000}(\vec{r})=\left(\frac{m\omega}{\hbar\pi}\right)^{3/4}\rme^{-m\omega r^2/\hbar}\left(\begin{array}{c}0\\ 1\end{array}\right)
\end{equation}
with corresponding eigenvalues $E_{00}^\pm=\pm mc^2$. We also remark that the Foldy-Wouthuysen Hamiltonian (\ref{HFW}) can be written as
\begin{equation}\label{HFW2}
  {\cal H}_{\rm FW} = mc^2\sqrt{1+\frac{2H_{\rm NR}}{mc^2}}\otimes\tau_3\,,
\end{equation}
a form already observed for other relativistic Hamiltonians exhibiting a SUSY \cite{Junker2020}.

\section{Green's function}\label{Section4}
The SUSY established for the KGO in the previous section also allows
us to study Green's function associated with the KGO Hamiltonian (\ref{HKG}). Following the general approach of \cite{Junker2020} Green's function being defined by
\begin{equation}\label{GKG}
  {\cal G}(z):= \frac{1}{{\cal H} -z}\,, \qquad z\in\mathbb{C}\backslash{\rm spec\,}{\cal H}\,,
\end{equation}
can be expressed in terms of the iterated Green's function ${\cal G}_{\rm I}$, that is,
\begin{equation}\label{GKG^2}
  {\cal G}(z)= ({\cal H} + z) {\cal G}_{\rm I}(z^2)\,, \qquad {\cal G}_{\rm I}(z^2):= \frac{1}{{\cal H}^2 -z^2}\,.
\end{equation}
Noting that ${\cal H}^2 = 2mc^2(H_{\rm NR} +mc^2/2)\otimes 1$, the iterated Green's function can be written in terms of the non-relativistic Green's function $G_{\rm NR}(\varepsilon):= (H_{\rm NR} - \varepsilon)^{-1}$ associated with $H_{\rm NR}$ as follows.
\begin{equation}\label{gKG^2a}
   {\cal G}_{\rm I}(z^2) = \frac{1}{2mc^2}\,G_{\rm NR}(\varepsilon)\otimes 1\,,\qquad \varepsilon := \frac{z^2}{2mc^2} -\frac{mc^2}{2}= 2mc^2\left(\left(\frac{z}{2mc^2}\right)^2-\frac{1}{4}\right)\,.
\end{equation}
Inserting this into above relation (\ref{GKG^2}) results in
\begin{equation}\label{GKG2}
  {\cal G}(z)=\frac{1}{2mc^2}\left(
  \begin{array}{cc}
    (H_{\rm NR}+mc^2 + z)\,G_{\rm NR}(\varepsilon) & H_{\rm NR}\,G_{\rm NR}(\varepsilon) \\
    -H_{\rm NR}\,G_{\rm NR}(\varepsilon) & -(H_{\rm NR}+mc^2 - z)\,G_{\rm NR}(\varepsilon)
  \end{array}
\right)\,.
\end{equation}
Using the defining relation $H_{\rm NR}\,G_{\rm NR}(\varepsilon)=\varepsilon \, G_{NR}(\varepsilon)$ together with the second relation in (\ref{gKG^2a}) leads us to the closed-form expression
\begin{equation}\label{GKGfinal}
  {\cal G}(z)=G_{\rm NR}(\varepsilon)\left(
  \begin{array}{cc}
    \left( \frac{1}{2}+\frac{z}{2mc^2}\right)\left(\frac{1}{2}+\frac{z}{2mc^2}\right)\, &
    \left( \frac{1}{2}+\frac{z}{2mc^2}\right)\left(\frac{z}{2mc^2}- \frac{1}{2}\right) \\[2mm]
    \left( \frac{1}{2}+\frac{z}{2mc^2}\right)\left(\frac{1}{2}-\frac{z}{2mc^2}\right) &
    \left( \frac{1}{2}-\frac{z}{2mc^2}\right)\left(\frac{z}{2mc^2}- \frac{1}{2}\right)
  \end{array}
\right)\,,
\end{equation}
with $\varepsilon$ as defined in (\ref{gKG^2a}). The reader is invited to verify that ${\cal H}\,{\cal G}(z) = z\, {\cal G}(z)$. With the definition $\tanh\vartheta := \varepsilon/(\varepsilon + mc^2)$ the above result may be put into the form
\begin{equation}\label{GKGfinal2}
  {\cal G}(z)=\frac{G_{\rm NR}(\varepsilon)}{\cosh\frac{\vartheta}{2}-\sinh\frac{\vartheta}{2}}\left(
  \begin{array}{cc}
    \cosh^2\frac{\vartheta}{2}\, & \cosh\frac{\vartheta}{2}\sinh\frac{\vartheta}{2}\\[2mm]
    -\cosh\frac{\vartheta}{2}\sinh\frac{\vartheta}{2} & -\sinh^2\frac{\vartheta}{2}
  \end{array}
\right)\,.
\end{equation}
To complete the discussion let us finally note that
the coordinate representation of ${G}_{\rm NR}(\vec{r},\vec{r}\,',\varepsilon):= \langle\vec{r}|{G}_{\rm NR}(\varepsilon)|\vec{r}\,'\rangle$ is known for long, see for example \cite{Blinder1984}, and explicitly reads
\begin{equation}\label{Grr'}
\begin{array}{l}
  \displaystyle
  {G}_{\rm NR}(\vec{r},\vec{r}\,',\varepsilon) = \frac{1}{rr'}\sum_{\ell = 0}^{\infty}{G}_{\ell}(r,r',\varepsilon)
       \sum_{\mu=-\ell}^{\ell}{\rm Y}^*_{\ell\mu}(\vec{e}\,'){\rm Y}_{\ell\mu}(\vec{e})\,, \\[2mm]
  \displaystyle
  {G}_{\ell}(r,r',\varepsilon)=-\frac{\Gamma\left(\frac{\ell}{2}-\frac{\varepsilon}{2\hbar\omega}\right)}{\sqrt{rr'}\,\hbar\omega}
  {\rm W}_{\lambda,\nu}\left(r^2_> m\omega/\hbar\right){\rm M}_{\lambda,\nu}\left(r^2_< m\omega/\hbar\right)\,,
\end{array}
\end{equation}
where ${\rm W}_{\lambda,\nu}$ and ${\rm M}_{\lambda,\nu}$ denote Whittaker's functions, and we have set $\lambda:= \frac{\varepsilon}{2\hbar\omega}+\frac{3}{4}$, $\nu:= \frac{\ell}{2}+\frac{1}{4}$, $r_>:=\max\{r,r'\}$ and $r_>:=\min\{r,r'\}$.
\section{Summary and~Outlook}\label{Section5}
In this work we have shown that the KGO exhibits a SUSY structure following closely the general approach of ref.\ \cite{Junker2020}. The SUSY of the KGO is found to be unbroken but, remarkably, with a vanishing Witten index. Despite the fact that the eigenvalues in (\ref{HPsi}) are known for a long time, see for example \cite{Debergh1992,Bruce1993}, the associated eigenstates in (\ref{HPsi}) have to our knowledge never been presented. Note that in \cite{Bruce1993} only the eigenstates of ${\cal H}_{\rm FW}$ are given. In addition, SUSY has enabled us to calculate the KGO Green function in a closed form.

Obviously, the current discussion for an isotropic oscillator may be extended to that for the anisotropic oscillator following Bruce and Manning \cite{Bruce1993}. Here, in essence, one needs to reduce the problem to three one-dimensional harmonic oscillators. An explicit expression for the Green's function may also be obtained as the one-dimensional harmonic oscillator Green's function is known in closed form, too. See, for example, Glasser and Nieto \cite{Glasser2015} and the discussion of the associated Dirac problem \cite{Junker2020b}. One may also pursue the path integral approach for the KGO along the lines of the corresponding approach for the Dirac oscillator \cite{Junker2018}. Another route for further investigation would be to look at generalised nonharmonic oscillators characterized by a potential function $U_a(r):= \lambda_a r^a$ using the minimal substitution $\vec{\pi}:=\vec{p}-\rmi (\vec{\nabla}U_a)(r)$. Such power-law potentials are known to obey a duality symmetry in classical and non-relativistic quantum mechanics, see the recent work \cite{Inomata2021} and references therein. In particular, the harmonic potential $a=2$, which corresponds to the discussed KGO case, is dual to the Kepler potential where $a=-1$.

Another extension, which might be entertained, is to apply the current SUSY construction to the relativistic $S=1$ oscillator. However, we must note that, as has been argued by Debergh et al \cite{Debergh1992}, the diagonal and off-diagonal matrix elements of the associated Hamiltonian no longer commute. Hence, it may not be possible to establish a SUSY structure.

\vspace{6pt}


\end{document}